# Improving Students' Understanding of Quantum Mechanics by Using Peer Instruction Tools

Chandralekha Singh and Guangtian Zhu

*Department of Physics and Astronomy, University of Pittsburgh, Pittsburgh, PA, 15260, USA*

**Abstract.** Quantum mechanics is a challenging subject, even for advanced undergraduate and graduate students. Here, we discuss the development and evaluation of research-based concept tests for peer instruction as a formative assessment tool in quantum mechanics (QM) courses. The preliminary evaluations show that these tools are effective in helping students develop a good grasp of quantum mechanics.

## INTRODUCTION

Development of research-based formative assessment tools is important at all levels of physics instruction. Peer instruction (PI), which can be used for formative assessment, was popularized by Mazur at Harvard University in the 1990s for introductory physics courses. As stated in Mazur's manual of PI, the fundamental goal of implementing peer instruction strategy in class is "to exploit student interaction during lectures and focus students' attention on underlying concepts" [1]. This statement points out two commonly existing problems in many physics classrooms. One problem is that students have little interaction with the instructor and their classmates during class, so they have inadequate opportunity to benefit from such interactions and reflect on what they are taught. Students are often too busy in taking notes to ask a question or discuss their confusions with the instructor and classmates. After class, they are very likely to forget about their questions. Some professors ask informal questions in the class to interact with the students. However, usually only a small group of students in the class are willing to answer the questions and the silent majority do not get involved.

The other common problem in traditional physics classes is that students pay less attention to qualitative interpretation compared with the quantitative skills they pick up while learning physics. One reason is that students only study what they are tested for. Since most of the questions in the homework and exams in a traditional physics course ask students to calculate a physical quantity or derive an equation, students easily arrive at the epistemological misunderstanding that physics is just a collection of formulas and algorithms. Without incentive, students make little effort to interpret the concepts and principles and learn to organize their knowledge hierarchically. They tend to use a plug-and-chug approach to solving physics problems by looking for a suitable formula in which they could make use of all the variables given in a problem statement. However, algorithmic exercises cannot improve students' conceptual understanding automatically. Research has shown that high-performing students on quantitative tests may fall in the low-performing group on conceptual tests [1]. Therefore, it is of great importance to help students develop conceptual understanding and help them build a good knowledge structure of physics.

When the peer-instruction approach is used for formative assessment, concept tests are used to lead peer discussions in class. A concept test question is usually a multiple-choice question related to a core concept or principle that is being discussed in the course. Most of the time, the options in each multiple-choice question have been prepared before the lecture (with alternative choices often dealing with common difficulties) though in some cases the instructor can ask the students to provide the possible answers and then let the class vote on these ideas. For a class using the peer instruction method, the class hour can be divided into several pieces of presentations focusing on each central point [2-3]. At the end of each short presentation, the corresponding concept test questions are given to the class. Students discuss with a partner the answers to the concept test questions and then they are polled either by electronic clickers, show of cards (with A through E written on each card) or by show of hands for each choice in the multiple-choice question.

## USE OF CONCEPT TESTS FOR QM

The unintuitive nature of QM implies that scaffolding is critical for helping students learn relevant concepts [4]. Scaffolding can be used to stretch students' learning far beyond their initial knowledge. Research-based formative assessment tools can be particularly helpful for providing scaffolding and continuous feedback as appropriate for students learning QM. We take into account the cognitive issues in learning QM and students' prior knowledge to develop the concept tests to help students build intuition about quantum phenomena and reduce their difficulties.

The development of concept tests goes through an iterative process to ensure that they are pedagogically valuable. The concept test questions related to a particular topic in QM are usually developed in a sequence to help students learn the same concept from different perspectives. Different representations of knowledge (verbal, graphical, mathematical etc.) are also exploited in a sequence of concept tests to help students develop a better grasp of the QM concepts. For example, we can use mathematical representation in one question and graphical representation in the next question to help students build intuition about the abstract concepts in QM. Many concept test questions are designed in sequences which are "easy-moderate-difficult" or "easy-difficult-difficult" types [5] to help students organize their knowledge better. Students are also given concept tests in which they discuss with their peers what should happen in a given situation and then they observe simulations to help them visualize the situation. For example, students are asked a sequence of four concept test questions dealing with the evolution of a wave function after the measurement of different physical observables before viewing simulations illustrating these concepts. These concept tests also help students understand the difference between the stationary state wave functions and position eigenfunctions, a topic about which students have many common misconceptions.

We have so far developed about 500 research-based concept test questions which are in the multiple-choice format. Similar to introductory physics, the QM concept test questions typically focus on the conceptual aspects. In some of the concept test questions, students are expected to have a basic knowledge of calculus and linear algebra. However, complicated mathematical manipulations are not involved in any concept tests.

Similar to the introductory physics courses, concept tests for QM can be integrated into lectures at 10-15 minute intervals or at the beginning of a lecture to reinforce material from the previous lecture. Posing research-based review questions at the beginning of the lectures ensures that students are the ones who do the thinking, organizing, repairing and extending of their knowledge structure.

When a concept test question is posed in the class, students must first consider the answer by themselves and then discuss it with their partners. Students are usually given one to two minutes to answer each concept-test question depending upon the complexity of the question. After the students submit their answers typically using clickers, the instructors discuss the correct answers with them and lead further discussion according to the distribution of students' answers. Since the concept tests are formative assessment tools and provide timely feedback to students, instructors can encourage students by awarding some credit to them for answering the questions and discussing them with peers even if they select the incorrect choice.

## BENEFITS OF CONCEPT TESTS

The following features of the peer instruction material and approach make it particularly suited for the challenging task of teaching QM: (1) Formative assessment by polling students about their responses provides feedback to the instructors which is critical for bridging the gap between teaching and learning. (2) The material is being developed based upon prior research by us and others on student difficulties and misconceptions related to QM. (3) The material strives to bridge the gap between the abstract quantitative formalism of QM and the qualitative understanding necessary to explain and predict diverse physical phenomena. (4) The method consistently keeps students actively engaged in the learning process because not only must the students answer the questions, they are encouraged to also discuss it with their peers. (5) The method provides a mechanism to continuously make students think about the material consistent with the goals of the course and the level of understanding that is desired of students. It can also help students monitor their own learning.

Similar to introductory physics, instant feedback on concept tests from students provides a "reality check" to the instructors teaching QM about the extent to which students have actually learned to apply the concepts discussed. This can help instructors adjust the pace of the class appropriately. Peer interaction also keeps students alert during the lectures because they are encouraged to discuss the questions with peers, and it also helps students organize and extend their knowledge. Articulating one's opinion requires attention to logic and organization of the thought process. Moreover, there is often a mismatch between the instructor and students' expectations about the level of understanding that is desired related to a concept. Peer instruction also helps convey the instructor's expectations explicitly and concretely to the students.

### A Sequence of Concept Test Questions

In the following section, we will discuss several concept test questions about the bound and scattering state wavefunctions which have been used in a junior-senior level QM course as formative assessment tools. The first concept-test question (CT1) is a relatively straightforward one which helps students reflect upon a basic model of a bound state. In a class with 18 students to which the question was posed, most students correctly recognized that the energy level given in CT1 (also see Figure 1) corresponds to a bound state for a 1D finite square well. Only 2 students incorrectly chose option A which represents an incorrect conception that the energy eigenfunction for a given quantum system can be a bound state and a scattering state simultaneously.

*Concept Test Question 1*
*Which one of the following statements is correct about an electron in a finite square well with a definite energy E as shown in Figure 1.*
A. *The electron is in a bound state between x=0 and x=a and is in a scattering state everywhere else.*
B. *The electron is in a bound state.*
C. *The electron is in a scattering state.*
D. *Whether the electron is in a bound or scattering state cannot be determined without knowing the wavefunction of the electron.*
E. *None of the above*

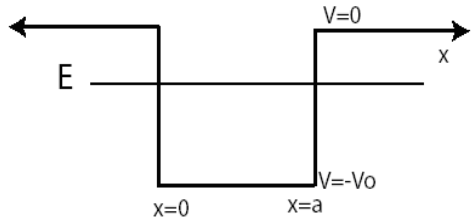

**Figure 1. An electron with energy $E$ interacting with a 1D finite square well.**

*Concept Test Question 2*
*Choose all of the following 1D potential energy functions that allow both bound and scattering states:*
(1) Finite square well
(2) Delta function potential energy well
(3) Delta function potential energy barrier
(4) Simple harmonic oscillator potential energy
A. *1 only*  B. *2 only*  C. *1 and 2 only*
D. *1, 2 and 3 only*  E. *all of the above*

The second concept test question (CT2) has a moderate difficulty level. Students are asked to review several potential energy functions and judge which function allows both bound and scattering states. About 72% of the students in a class of 18 students chose the correct models, i.e., the 1D finite square potential energy well and the 1D delta function potential energy well. After a brief discussion of the different functions in CT2, the instructor asked students a difficult question as the third concept test question (CT3) of the sequence.

*Concept Test Question 3*
*Choose all of the following 1D potential energies (Figure 2) that allow(s) both bound and scattering states.*
A. *all*  B. *2 only*  C. *2 and 4 only*  D. *3 and 4 only*
E. *2, 3 and 4 only*

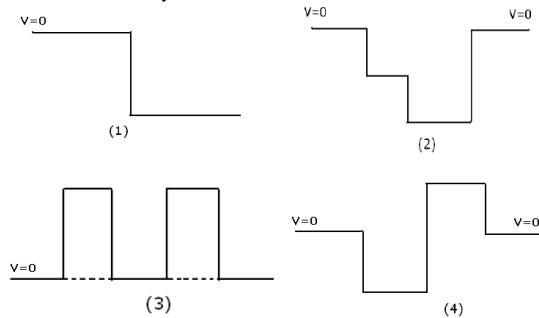

**Figure 2. Potential energies that may allow bound states or scattering states or both.**

The third concept test question (CT3) asks students to choose the potential energies that allow both bound and scattering states. However, unlike the verbal representation of the basic models in CT2, graphical representation is used to illustrate the different models of potential energy wells in CT3. Since the students had not explicitly learned in detail about these potential energies before, they must apply their previous knowledge about bound and scattering states to these novel situations in CT3. Only 33% of the students (out of 18) chose the correct answer C and 50% of them incorrectly believed that option (3) also allows both bound and scattering state wavefunctions. In fact, only scattering state wavefunctions can exist for the potential energy in option (3) because the possible energy levels are always higher than the potential energy at plus and minus infinity. Students had an active discussion before they submitted the answer for CT3. When some students found that their ideas were incorrect, they were eager to resolve the discrepancy between their answers and the correct away to conceptualize the ideas applicable in CT3.

## PRELIMINARY EVALUATION

We have administered several tests and surveys to investigate the effectiveness of research-based concept tests as formative assessment tools for learning QM [6]. We find that students' understanding of QM improved significantly after using concept tests as peer instruction tools. To illustrate the improvement in students' performance after the concept tests, here we discuss the results of a quiz about the 1D infinite square potential energy well which is designed to examine students' understanding of some basic concepts in QM. The quiz on the 1D infinite square well was administered to three classes (experimental groups) taking a junior-senior level QM course with concept tests in 2008, 2009 and 2010. The numbers of students in these classes were 25, 13 and 20, respectively. The concept test questions used in the three experimental groups were not exactly the same because we kept refining the concept tests based on both the instructors' suggestions and students' responses from the previous years. We also gave the quiz to a comparison group of 18 students receiving traditional instruction in QM who did not use concept tests. There is a significant difference ($p<0.0001$) in students' performances between the experimental groups (with lectures and concept tests) and the comparison group with only traditional lectures. We have also observed continuous improvement in the experimental groups' performance from 2008 to 2010 as we refined the concept tests based upon the feedback obtained.

The quiz contained 7 multiple-choice questions and 3 open-ended questions, all of which tested the basic quantum-mechanical concepts related to the model of a 1D infinite square potential energy well, e.g., possible wavefunctions allowed in an infinite square well, time evolution of the wavefunction in the well, energy or position measurement, etc. The total score for this quiz is 10 and the average scores for the three experimental groups in 2008, 2009 and 2010 are 5.5, 7.0 and 7.6, respectively. The average score for the comparison group students is only 1.8. The distribution of students' individual scores is shown in Figure 3. The horizontal axis represents the possible scores that a student can obtain and the vertical axis represents the percentage of students in each of the four groups who obtained a particular score. The comparison group and the experimental groups in 2008, 2009 and 2010 are marked in the different shaded patterns as shown in the legend. For example, the lightest bar on the rightmost side of Figure 3 indicates that 30% of the students in the experimental group in 2010 obtained the full score (10 points) on the quiz. In the comparison group with only traditional instruction, most students obtained scores below 3 points and only one student obtained 6 points (the highest score for this group). However, in each of the experimental groups with concept tests, most students obtained scores greater than 4 points. Figure 3 shows that as we refined the concept tests from 2008 to 2010, higher percentages of students obtained high scores (9 or 10 points).

## SUMMARY

We have used concept tests as a peer instruction tool for formative assessment in junior-senior level QM courses for three consecutive years. The comparison between the classes using concept tests and the class having only traditional lectures suggests that the concept tests are effective in improving students' understanding of QM. We also observed continuous improvement in students' performance on the conceptual surveys of QM as we refined the concept test questions from 2008 to 2010.

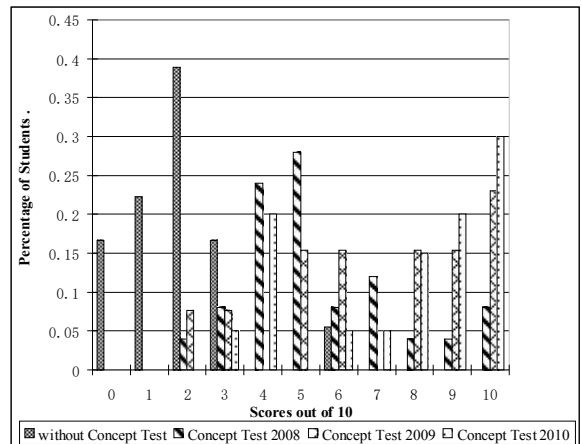

**Figure 3.** The distribution of students' individual scores on the 1D infinite square well quiz. The horizontal axis represents the possible scores on the quiz that a student can obtain and the vertical axis represents the percentage of students in each of the four groups who obtained a particular score. The comparison group and the experimental groups in 2008, 2009 and 2010 are marked with the different shaded patterns as shown in the legend.

## ACKNOWLEDGMENTS

We thank the National Science Foundation for awards PHY-0968891 and PHY-0653129.